\documentstyle[preprint,tighten,eqsecnum,epsf,aps]{revtex}
\begin{document}
\draft

\title{Nonclassical interferometry with intelligent light}
\author{C. Brif \ \cite{email1} \ and \ A. Mann \cite{email2}}
\address{Department of Physics, Technion -- Israel Institute of 
Technology, Haifa 32000, Israel}
   \maketitle

	\begin{abstract}
We study the phase sensitivity of SU(2) and SU(1,1) interferometers 
fed by two-mode field states which are intelligent states for 
Hermitian generators of the SU(2) and SU(1,1) groups, respectively. 
Intelligent states minimize uncertainty relations and this makes 
possible an essential reduction of the quantum noise in 
interferometers. Exact closed expressions for the minimum detectable 
phase shift are obtained in terms of the Jacobi polynomials. These 
expressions are compared with results for some conventional input 
states, and some known results for the squeezed input states are 
reviewed. It is shown that the phase sensitivity for an 
interferometer that employs squeezing-producing active devices (such 
as four-wave mixers) should be analyzed in two regimes: (i) fixed 
input state and variable interferometer, and (ii) fixed 
interferometer and variable input state.
The behavior of the phase sensitivity is essentially different in 
these two regimes. The use of the SU(2) intelligent states allows us
to achieve a phase sensitivity of order $1/\bar{N}$ (where $\bar{N}$
is the total number of photons passing through the phase shifters of 
the interferometer) without adding four-wave mixers. This avoids the 
duality in the behavior of the phase sensitivity that occurs for the 
squeezed input. On the other hand, the SU(1,1) intelligent states have
the property of achieving the phase sensitivity of order $1/\bar{N}$ 
in both regimes.
	\end{abstract}


\section{Introduction}
\label{sec:intr}

A lot of attention has been recently paid to the improvement of 
measurement accuracy of interferometers, because this problem is of 
great importance in many areas of experimental physics. A very 
promising way to reduce quantum fluctuations in interferometers is 
based on the use of input light fields prepared in special quantum 
states. Therefore, with further development of technology, 
high-accuracy interferometry seems to have become one of the most 
important applications of nonclassical photon states whose properties 
are now extensively studied in the quantum optics literature. 

The first steps in this area were taken by Caves \cite{Caves} and 
Bondurant and Shapiro \cite{BoSh} who showed that the use of 
squeezed light can reduce the quantum 
noise in interferometers \cite{Paris}. Yurke, McCall and Klauder 
\cite{YMK} used powerful group-theoretic methods for the study of 
interferometers employing passive and active optical devices.
The interferometers considered in \cite{Caves,BoSh} employ passive 
lossless devices, such as beam splitters. Yurke, McCall and Klauder 
\cite{YMK} showed that such interferometers can be characterized by 
the SU(2) group. They also introduced a class of interferometers 
which employ active lossless devices, such as four-wave mixers, and 
are characterized by the SU(1,1) group. 
The actual problem of high-accuracy interferometry is the improvement 
of the phase sensitivity, i.e., the optimization of the minimum 
detectable phase shift $\delta\phi$ for a given mean total number 
$\bar{N}$ of photons passing through phase shifters. This problem 
arises because of the back-action effect of the radiation pressure. 
It was shown \cite{YMK} that SU(2) interferometers can achieve a phase 
sensitivity $\delta\phi \sim 1/\bar{N}$ provided that light entering 
the input ports is prepared in a two-mode squeezed state. SU(1,1) 
interferometers can achieve this sensitivity even when the vacuum 
fluctuations enter the input ports \cite{YMK}. Holland and Burnett
\cite{HoBu} have considered the reduction of the uncertainty in the
relative quantum phase of two field modes propagating in an SU(2) 
interferometer fed by two Fock states with equal numbers of photons.
They considered \cite{HoBu} the specific ``reduced'' situation of the
measurement with the sensitivity measure different from that used in
Ref.\ \cite{YMK}.

In a separate line of research, considerable efforts have been 
devoted during the last few years to generalize the idea of squeezing
to the SU(2) and SU(1,1) Lie groups. The usual squeezed states are 
the generalized coherent states of SU(1,1) \cite{Per}, i.e., they are 
produced by the action of the group elements on the extreme state of 
the group representation Hilbert space. Another interesting class of 
states which has been considered is the class of the so-called 
intelligent states \cite{Arag}, which minimize the uncertainty 
relations for the Hermitian generators of the group. Squeezing 
properties of the SU(2) and SU(1,1) intelligent states have been 
widely discussed in the literature 
\cite{WE,Ger85_88,Hil87_89,Agar88,AgPu,Buz90,NiTr,BHY_YH,PrAg94,Trif,%
BBA:jpa,GeGr,PrAg95}.
Recently, Nieto and Truax \cite{NiTr} showed that a generalization
of squeezed states for an arbitrary dynamical symmetry group leads to 
the intelligent states for the group generators. Connections between
the concepts of squeezing and intelligence were further investigated 
by Trifonov \cite{Trif}. It turns out that the intelligent states for
two Hermitian operators can provide an arbitrarily strong squeezing 
in either of these observables \cite{Trif}. Some schemes for the 
experimental production of the SU(2) and SU(1,1) intelligent states 
in nonlinear optical processes have been suggested recently by a 
number of authors \cite{AgPu,PrAg94,GeGr,LuisPer}. The most recent
scheme, developed by Luis and Pe\v{r}ina \cite{LuisPer}, is of 
remarkable physical elegance and conceptual clarity and seems to be 
technically realizable.

The group-theoretic analysis of interferometers and the 
group-theoretic generalization of squeezing (i.e., intelligence)
were brought together by Hillery and Mlodinow \cite{HM} who proposed
to use intelligent states of the two-mode light field for increasing
the precision of interferometric measurements. They derived \cite{HM}
approximate results for the phase sensitivity of an SU(2) 
interferometer fed with the SU(2) intelligent states. The possibility
to improve further the accuracy of SU(1,1) interferometers by using 
specially prepared input states has been also studied recently 
\cite{BBA:qso}. It was shown \cite{BBA:qso} that the use of 
two-mode SU(1,1) coherent states which are simultaneously the SU(1,1) 
intelligent states can improve the measurement accuracy when the 
photon-number difference between the modes is large. 

In the present work we consider in detail both SU(2) and SU(1,1)
interferometers whose input ports are fed with intelligent light.
We use powerful analytic methods that employ representations 
of intelligent states in the generalized coherent-state bases.
Thus we are able to obtain exact analytic expressions for the
phase sensitivity and examine them in various limits. These results
are compared with those obtained in the cases when the input
field is prepared in the usual coherent state, in the generalized 
coherent state and in the squeezed state. We show that the use of
squeezing-producing active devices (such as four-wave mixers)
introduces a duality in the behavior of the phase sensitivity.
For example, when the squeezed input states are used, the 
interferometer can be operated in two regimes: with variable
squeezing parameter and fixed coherent amplitude, and vice versa.
The regime of variable squeezing leads to the phase sensitivity
$\delta\phi \sim 1/\bar{N}$, whereas the technically preferable 
regime of variable coherent amplitude gives only $\delta\phi \sim 
1/\bar{N}^{1/2}$ (the standard noise limit). The use of the SU(2)
intelligent states avoids this dual behavior and leads to the
phase sensitivity $\delta\phi \sim 1/\bar{N}$ without adding a
four-wave mixer to the interferometer. The SU(1,1) intelligent 
states also allow us to obtain a significant improvement of the 
measurement accuracy. These states exhibit a very specific behavior
providing phase sensitivity of order $1/\bar{N}$ in the two
regimes: variable interferometer, and variable input state.
We emphasize that the optimization of the phase sensitivity by 
the intelligent input states is a consequence of their remarkable 
squeezing properties. 

\section{SU(2) interferometers with conventional input states}
\label{sec:su2_reg}

\subsection{The interferometer}
\label{ssec:su2_scheme}

An SU(2) interferometer is described schematically in 
Fig.\ 1. Two light beams represented by the mode annihilation 
operators $a_{1}$ and $a_{2}$ enter the first beam splitter BS1. 
After leaving BS1, the beams accumulate phase shifts $\phi_{1}$ and 
$\phi_{2}$, respectively, and then they enter the second beam 
splitter BS2. The photons leaving the interferometer are counted by 
detectors D1 and D2.

For the analysis of such an interferometer it is convenient to 
consider the Hermitian operators
\begin{eqnarray}
& & J_{1} = \frac{1}{2}(a_{1}^{\dagger}a_{2} + 
a_{2}^{\dagger}a_{1}) , \nonumber \\
& & J_{2} = \frac{1}{2i}(a_{1}^{\dagger}a_{2} - 
a_{2}^{\dagger}a_{1}) , \label{2.1} \\
& & J_{3} = \frac{1}{2}(a_{1}^{\dagger}a_{1} - 
a_{2}^{\dagger}a_{2}) .   \nonumber 
\end{eqnarray}
These operators form the two-mode boson realization of the SU(2)
Lie algebra:
\begin{eqnarray}
& &  [ J_{1}, J_{2} ] = iJ_{3}  , \nonumber \\   
& &  [ J_{2}, J_{3} ] = iJ_{1}  , \label{2.2} \\
& &  [ J_{3}, J_{1} ] = iJ_{2}  .  \nonumber
\end{eqnarray}
It is also useful to introduce the raising and lowering operators
\begin{eqnarray}
& & J_{+} = J_{1}+iJ_{2} = a_{1}^{\dagger}a_{2} , \nonumber \\ 
& & J_{-} = J_{1}-iJ_{2} = a_{2}^{\dagger}a_{1} .
\label{2.3} 
\end{eqnarray}
The Casimir operator for any unitary irreducible representation of 
SU(2) is a constant
\begin{equation}
J^{2} = J_{1}^{2} + J_{2}^{2} + J_{3}^{2} = j(j+1) ,  \label{2.4}
\end{equation}
and a representation of SU(2) is determined by a single number
$j$ that acquires discrete positive values 
$j=\frac{1}{2},1,\frac{3}{2},2,\ldots$.
By using the operators of Eq.\ (\ref{2.1}), one gets
\begin{equation}
J^{2} = \frac{N}{2} \left( \frac{N}{2} +1 \right)  , 
\label{2.5}  \end{equation}
where
\begin{equation}
N = a_{1}^{\dagger}a_{1} + a_{2}^{\dagger}a_{2}   
\label{2.6}  \end{equation}
is the total number of photons entering the interferometer. We see
that $N$ is an SU(2) invariant related to the index $j$ via
$j = N/2$. The representation Hilbert space is spanned by the 
complete orthonormal basis $|j,m\rangle$ ($m=-j,-j+1,\ldots,j-1,j$)
that can be expressed in terms of Fock states of two modes:
\begin{equation}
|j,m\rangle = |j+m\rangle_{1} |j-m\rangle_{2} .
\label{2.7}  \end{equation}

The actions of the interferometer elements on the vector
$\bbox{J}=(J_{1},J_{2},J_{3})$ can be represented as rotations in 
the 3-dimensional space \cite{YMK}. BS1 acts on $\bbox{J}$ as a 
rotation about the 1st axis by the angle $\pi/2$. 
The transformation matrix of this rotation is 
\begin{equation}
{\sf R}_{1}(\pi/2) = \left( \begin{array}{ccc}
1 & 0 & 0 \\ 0 & 0 & -1 \\ 0 & 1 & 0 \end{array} \right) .
\label{2.8}  \end{equation}
The transformation matrix of BS2 is ${\sf R}_{1}(-\pi/2)$, i.e., 
the two beam splitters perform rotations in opposite directions. 
The phase shifters rotate $\bbox{J}$ about the 3rd axis by an angle 
$\phi=\phi_{2}-\phi_{1}$. The transformation matrix of this 
rotation is 
\begin{equation}
{\sf R}_{3}(\phi) = \left( \begin{array}{ccc}
\cos\phi & -\sin\phi & 0 \\ \sin\phi & \cos\phi & 0 \\ 0 & 0 & 1 
\end{array} \right) .
\label{2.9}  \end{equation}
The overall transformation performed on $\bbox{J}$ is
\begin{equation}
\bbox{J}_{{\rm out}} = 
{\sf R}_{1}(-\pi/2){\sf R}_{3}(\phi){\sf R}_{1}(\pi/2) \bbox{J} .
\label{2.10}  \end{equation}

The information on the phase shift $\phi$ is inferred from the 
photon statistics of the output beams. One should measure the  
difference between the number of photons in the two output modes, 
$(N_{d})_{{\rm out}}$, or, equivalently, the operator 
$J_{3\,{\rm out}} = \frac{1}{2}(N_{d})_{{\rm out}}$.
Since there are fluctuations in $J_{3\,{\rm out}}$, a phase shift
is detectable only if it induces a change in 
$\langle J_{3\,{\rm out}} \rangle$ which is larger than $\Delta
J_{3\,{\rm out}}$. Therefore, the minimum detectable phase shift
(i.e., the uncertainty of the phase measurement) is determined by  
\begin{equation}
(\delta\phi)^{2} = \frac{ (\Delta J_{3\,{\rm out}})^{2} }{ \left|
\partial\langle J_{3\,{\rm out}}\rangle /\partial\phi \right|^{2}} .
\label{2.11}  \end{equation}
The value of $\delta\phi$ characterizes the accuracy of 
the interferometer. The expression for $J_{3\,{\rm out}}$ can be 
easily found by using Eq.\ (\ref{2.10}):
\begin{equation}
J_{3\,{\rm out}} = -(\sin\phi)J_{1} + (\cos\phi)J_{3} .   
\label{2.12}  \end{equation}

\subsection{The standard noise limit}
\label{ssec:snl}

We consider some typical input states for which the phase sensitivity
of an SU(2) interferometer is restricted by the so-called standard
noise limit (SNL). 
Let the input state be $|j,m\rangle = |j+m\rangle_{1} |j-m\rangle_{2}$ 
(an eigenstate of $J_3$ with eigenvalue $m$). The phase sensitivity
for this input state is obtained from Eq.\ (\ref{2.11}) by a
straightforward calculation:
\begin{equation}
(\delta\phi)^{2} = \frac{j^2 - m^2 + j}{ 2 m^2 } , 
\mbox{\hspace{0.4cm}} \phi \neq 0\; ({\rm mod}\, \pi) .  \label{jmsens}
\end{equation}
In this situation the best phase sensitivity is obtained for
$m = \pm j$. Thus for the input state $|j,j\rangle = 
|2j\rangle_{1}|0\rangle_{2}$, one gets \cite{YMK}
\begin{equation}
(\delta\phi)^{2}_{\rm SNL} = 1/(2j) = 1/N , 
\mbox{\hspace{0.4cm}} \phi \neq 0\; ({\rm mod}\, \pi) .  \label{2.13}
\end{equation}
This means that the phase sensitivity $\delta\phi$ of the 
interferometer goes as $1/\sqrt{N}$. The phase sensitivity 
(\ref{2.13}) is usually referred to as the standard noise limit
\cite{YMK}.

It follows from Eq.\ (\ref{jmsens}) that for the input
state $|j,m\rangle$ with $m=0$ (i.e., when the interferometer is
fed by two Fock states with equal numbers of photons), the phase
measurement is absolutely uncertain [under the condition 
$\phi \neq 0\; ({\rm mod}\, \pi)$]. This result is in accordance
with qualitative arguments of Yurke, McCall and Klauder (see Fig.\ 
2 of Ref.\ \cite{YMK}). However, it has been shown by Holland and 
Burnett \cite{HoBu} that this input state can be used in an SU(2)
interferometer with the specific ``reduced'' situation of the 
measurement of the relative quantum phase between two field modes.
In the Holland-Burnett situation the use of the simplified 
sensitivity measure (\ref{2.11}) is excluded. 

In what follows we assume, for the sake of simplicity, $\phi = 0$. 
This can be achieved by controlling $\phi_{2}$ with a feedback loop 
which maintains $\phi = \phi_{2}-\phi_{1} = 0$ \cite{YMK}. Then Eq.\ 
(\ref{2.11}) with $J_{3\,{\rm out}}$ given by (\ref{2.12}) can be
simplified to the form
\begin{equation}
(\delta\phi)^{2} = \frac{ (\Delta J_{3})^{2} }{
\langle J_{1}\rangle^{2}} , 
\mbox{\hspace{0.4cm}} \langle J_{1}\rangle \neq 0 .    
\label{2.14} 
\end{equation}
Consider now the input state $|\alpha\rangle_1 |\alpha'\rangle_2$,
where
\begin{equation}
|\alpha\rangle = \exp(-|\alpha|^2/2) \sum_{n=0}^{\infty}
\frac{\alpha^n}{\sqrt{n!}} |n\rangle
\end{equation}
is the familiar Glauber coherent state. A simple calculation yields
\begin{eqnarray} 
& & (\Delta J_{3})^{2} = ( |\alpha|^2 + |\alpha'|^2 )/4 , \\
& & \langle J_{1} \rangle = |\alpha| |\alpha'| 
\cos(\theta + \theta') ,
\end{eqnarray}
where $\alpha = |\alpha|\, e^{i\theta}$, 
$\alpha' = |\alpha'|\, e^{i\theta'}$.
For the optimal choice $\theta + \theta' = 0$, we get
\begin{equation}
(\delta\phi)^{2} = \frac{ |\alpha|^2 + |\alpha'|^2 }{ 4 
|\alpha|^2 |\alpha'|^2 } .
\end{equation}
The total number of photons is $N=|\alpha|^2 + |\alpha'|^2$. Hence
the best phase sensitivity is obtained for 
$|\alpha|^2 = |\alpha'|^2 = N/2$ and it achieves the standard noise 
limit of Eq.\ (\ref{2.13}).

We also consider the SU(2) generalized coherent states that are 
defined by \cite{Per}
\begin{eqnarray}
|j,\zeta\rangle & = & \exp(\xi J_{+}-\xi^{\ast}J_{-}) |j,-j\rangle 
= \frac{ \exp(\zeta J_{+}) }{ (1+|\zeta|^{2})^{j} } |j,-j\rangle 
\nonumber \\ & = & (1+|\zeta|^{2})^{-j} 
\sum_{m=-j}^{j} \left[ \frac{(2j)! }{ (j+m)!(j-m)! } \right]^{1/2} 
\zeta^{j+m} |j,m\rangle ,  \label{2.15}     
\end{eqnarray}
where $\zeta = (\xi/|\xi|)\tan|\xi|$. Expectation values of the
SU(2) generators can be easily calculated for the $|j,\zeta\rangle$
states:
\begin{eqnarray} 
& & (\Delta J_{3})^{2} = 2j|\zeta|^{2}/(1+|\zeta|^{2})^{2} ,
\label{2.16} \\ & & \langle J_{1} \rangle = 
2j ({\rm Re}\,\zeta)/(1+|\zeta|^{2}) .  \label{2.17}
\end{eqnarray}
Then Eq.\ (\ref{2.14}) reads 
\begin{equation}
(\delta\phi)^{2}_{{\rm coh}}=\frac{|\zeta|^{2}}{ 2j
({\rm Re}\,\zeta)^{2} } . 
\label{2.18}         \end{equation}
This phase uncertainty is minimized when $\zeta$ is real. Then 
$(\delta\phi)^{2}_{{\rm coh}}$ achieves the standard noise limit 
of Eq.\ (\ref{2.13}). We see that the use of the Glauber coherent 
states and of the SU(2) generalized coherent states does not improve 
the measurement accuracy over the standard noise limit. 

\subsection{Squeezed input states, the role of active devices
and a duality of the phase sensitivity}
\label{ssec:squ}

There have been attempts to surpass the standard noise limit by
using squeezed input states \cite{Caves,BoSh,YMK}. We reconsider
here the scheme proposed by Yurke, McCall and Klauder \cite{YMK}.
They considered the SU(2) interferometer outlined in Fig.\ 1 whose
input ports are fed by the output beams $b_1$ and $b_2$ of a four-wave
mixer (see Fig. 5 of Ref.\ \cite{YMK}). The transformation caused by
the four-wave mixer on the light beams $a_1$ and $a_2$ entering its 
input ports is an SU(1,1) transformation \cite{YMK,Leon}:
\begin{equation}
\left( \begin{array}{c} b_1 \\ b_2^{\dagger} \end{array} \right) =
\left( \begin{array}{cc} \cosh(\beta/2) & \sinh(\beta/2) \\
\sinh(\beta/2) & \cosh(\beta/2) \end{array} \right)
\left( \begin{array}{c} a_1 \\ a_2^{\dagger} \end{array} \right) .
\label{trans}
\end{equation}
The parameter $\beta$ is related to the reflectivity $r$ of the 
four-wave mixer (when it is used as a phase-conjugating mirror) via
$\sinh^2 (\beta/2) = r$ \cite{ReWa}. In the scheme considered here
the Glauber coherent state $|\alpha\rangle$ enters one input port of
the four-wave mixer and the vacuum state $|0\rangle$ enters the other.
Since the transformation (\ref{trans}) is a squeezing Bogoliubov
transformation, the output state of the four-wave mixer is the 
two-mode squeezed state. 

The generator $J_3$ representing the photon-number difference between
the two modes is invariant under the transformation (\ref{trans}). 
Therefore one finds 
\begin{equation}
(\Delta J_{3})^{2} = |\alpha|^2 / 4 .
\end{equation}
The generator $J_1$ at the output of the four-wave mixer is given by
\begin{eqnarray}
J_1 & = & \mbox{$\frac{1}{4}$} \sinh\beta ( a_1^{\dagger 2} + a_1^2
+ a_2^{\dagger 2} + a_2^2 ) - \frac{i}{4} \sinh\beta
( a_1^{\dagger 2} - a_1^2 + a_2^{\dagger 2} - a_2^2 ) \nonumber \\
& & + \mbox{$\frac{1}{2}$} \cosh\beta 
( a_1^{\dagger} a_2 + a_2^{\dagger} a_1 ) .
\end{eqnarray}
Its expectation value for the input state $|\alpha\rangle_1 
|0\rangle_2$ is
\begin{equation}
\langle J_1 \rangle = \frac{1}{2} |\alpha|^2 \sinh\beta \cos 2\theta
\end{equation}
where $\alpha = |\alpha|\, e^{i\theta}$. The phase uncertainty of 
Eq.\ (\ref{2.14}) is minimized when $\theta = 0$. Then one obtains
\cite{YMK}
\begin{equation}
(\delta\phi)^{2}_{\alpha,\beta} = \frac{1}{|\alpha|^2 \sinh^2 \beta} .
\label{vz1}
\end{equation}

The measurement accuracy can be improved in two ways: (i) by
increasing the parameter $\beta$ of the four-wave mixer, or (ii)
by increasing the coherent-state intensity $|\alpha|^2$. The first way
can be viewed as related to the interferometer (including the four-wave
mixer), while the second is related to the input state. Therefore,
when we consider the phase sensitivity $\delta\phi(N)$, we should
distinguish between the sensitivity for fixed input state 
$(\alpha = {\rm const})$ and the sensitivity for fixed 
interferometer $(\beta = {\rm const})$. This distinction seems formal
at first look, but it has a crucial physical importance for an 
interferometer employing active devices because they do not conserve
the total number of photons. Indeed, when the four-wave mixer is 
applied, the total number of photons is not constant any more. The
mean total number $\bar{N}$ of photons passing through the phase 
shifters depends on both $\alpha$ and $\beta$. In the scheme presented 
here $\bar{N}$ is the mean total number of photons emitted by the
four-wave mixer:
\begin{equation}
\bar{N} = \langle b_1^{\dagger} b_1 +  b_2^{\dagger} b_2 \rangle
= (|\alpha|^2 + 1) \cosh\beta - 1 .   \label{vz2}
\end{equation}
Then we find the phase sensitivity for fixed input state:
\begin{equation}
\left. (\delta\phi)^{2} \right|_{\alpha} = \frac{ (|\alpha|^2 + 1)^2
}{ |\alpha|^2 [ (\bar{N}+1)^2 - (|\alpha|^2 + 1)^2 ] } , \label{1i}
\end{equation}
and for fixed interferometer:
\begin{equation}
\left. (\delta\phi)^{2} \right|_{\beta} = \frac{ \cosh\beta}{
\sinh^2 \beta} \frac{1}{ (\bar{N}+1-\cosh\beta) } .  \label{2i}
\end{equation}
When $|\alpha|^2$ is close to 1 and $\bar{N}$ is large, 
Eq.\ (\ref{1i}) yields
\begin{equation}
\left. \delta\phi \right|_{\alpha} \approx \frac{2}{\bar{N}} .
\end{equation}
This is much better than the standard noise limit, but there is a
subtlety. Actually, for $|\alpha|^2 \sim 1$ the range of $\bar{N}$
is restricted by available four-wave mixers. It is much more 
convenient for the experimenter to improve the measurement accuracy
by increasing the intensity of the coherent state $|\alpha\rangle$.
However, Eq.\ (\ref{2i}) shows that in this regime the standard 
noise limit cannot be surpassed. Therefore, when speaking about the
phase sensitivity achieved with the squeezed input states, it is 
necessary to specify the regime of operation of the interferometr.

\section{SU(2) interferometers with intelligent input states}
\label{sec:su2_int}

\subsection{The SU(2) intelligent states}
\label{ssec:su2_is}

It is known \cite{HM} that the standard noise limit for SU(2)
interferometers can be surpassed by using the SU(2) intelligent
states. However, an expression for $\delta\phi$ was found in Ref.\
\cite{HM} only for a special limiting case. We would like to derive
an exact analytic expression for $\delta\phi$, that holds for a wide
class of the SU(2) intelligent states. The commutation relation
$[J_{2},J_{3}]=iJ_{1}$ implies the uncertainty relation
\begin{equation}
(\Delta J_{2})^{2} (\Delta J_{3})^{2} \geq \frac{1}{4} 
\langle J_{1} \rangle^{2}  .       \label{2.19} 
\end{equation} 
Therefore, Eq.\ (\ref{2.14}) reads 
\begin{equation}
(\delta\phi)^{2} \geq \frac{1}{4(\Delta J_{2})^{2}} .
\label{2.20}  \end{equation}
For intelligent states an equality is achieved in the uncertainty 
relation. Such $J_{2}$-$J_{3}$ intelligent states with large values
of $\Delta J_{2}$ would allow us to measure small changes in $\phi$. 
The $J_{2}$-$J_{3}$ intelligent states $|\lambda,\eta\rangle$ are 
determined by the eigenvalue equation 
\begin{equation}
(\eta J_{2} + i J_{3}) |\lambda,\eta\rangle = \lambda 
|\lambda,\eta\rangle ,     \label{2.21}  
\end{equation}
where $\lambda$ is a complex eigenvalue and $\eta$ is a real 
parameter given by $|\eta| = \Delta J_{3}/ \Delta J_{2}$.
For $|\eta| > 1$, these states are squeezed in $J_{2}$, and for 
$|\eta| < 1$, they are squeezed in $J_{3}$. In what follows we
will consider only the region $|\eta| < 1$, that guarantees, as we
will see, an improvement of the measurement accuracy. The states
of Eq.\ (\ref{2.21}) can be generated from the vacuum in two
parametric down-conversion crystals with aligned idler beams
after a measurement of the photon number in some of the modes
\cite{LuisPer}.
For the $J_{2}$-$J_{3}$ intelligent states, Eq.\ (\ref{2.20}) reads
\begin{equation}
(\delta\phi)^{2}_{{\rm int}} = \frac{1}{ 4(\Delta J_{2})^{2} } 
= \frac{\eta^{2}}{ 4(\Delta J_{3})^{2} } .   \label{2.22}   
\end{equation}

Our aim is now to evaluate the variance $(\Delta J_{3})^{2}$. In
order to do that, we use the analytic representation of the
intelligent states in the coherent-state basis $|j,\zeta\rangle$.
This basis is overcomplete and any state in the Hilbert space can be
expanded in it \cite{Per}. For example, the SU(2) intelligent state
\begin{equation}
|\lambda,\eta\rangle = \sum_{m=-j}^{j} C_{m} |j,m\rangle 
\label{2.23}    \end{equation}
is represented by the entire analytic function
\begin{equation}
\Lambda(j,\lambda,\eta;\zeta) = (1+|\zeta|^2)^{j}\langle
j,\zeta^{\ast}|\lambda,\eta\rangle = \sum_{m=-j}^{j} C_{m} 
\left[ \frac{ (2j)! }{ (j+m)!(j-m)! } \right]^{1/2} \zeta^{j+m} .
\label{2.24}   \end{equation}
The SU(2) generators act on $\Lambda(\zeta)$ as first-order 
differential operators \cite{Per}:
\begin{equation}
J_{+} = -\zeta^{2} \frac{d}{d\zeta} + 2j\zeta ,  
\mbox{\hspace{0.4cm}} J_{-} = \frac{d}{d\zeta} ,
\mbox{\hspace{0.4cm}} J_{3} = \zeta \frac{d}{d\zeta} -j .    
\end{equation}
Then Eq.\ (\ref{2.21}) can be converted into a first-order linear 
homogeneous differential equation for $\Lambda(\zeta)$:
\begin{equation}
(\eta + 2\zeta + \eta \zeta^{2}) \frac{d\Lambda}{d\zeta} 
+ 2 (i\lambda -j -j\eta\zeta) \Lambda = 0 .
\end{equation}
The solution of this equation can be easily found to be
\begin{equation}
\Lambda(j,m_{0},\eta;\zeta) = {\cal N}^{-1/2}
(1+\zeta/\tau)^{j+m_{0}}(1+\tau\zeta)^{j-m_{0}} ,
\label{2.25}    \end{equation}
where ${\cal N}$ is a normalization factor, and we have defined
\begin{eqnarray}
& & \tau \equiv \left(1 - \sqrt{1-\eta^{2}} \right)/\eta ,  
\label{2.26}  \\ 
& & \lambda(m_{0}) \equiv i m_{0}\sqrt{1-\eta^{2}} . \label{2.27}
\end{eqnarray}
The analyticity condition for the function $\Lambda(\zeta)$ requires
that $m_{0}$ can take only the values:
\begin{equation}
m_{0} = -j,-j+1,\ldots,j-1,j .   \label{2.28}
\end{equation}
Then Eq.\ (\ref{2.27}) becomes a quantization condition which 
means that the operator $\eta J_{2} + i J_{3}$ has a discrete 
spectrum, and the corresponding eigenstates and eigenvalues are
characterized by the quantum number $m_{0}$. 

In the special cases
$m_{0} = \pm j$, the $J_{2}$-$J_{3}$ intelligent states 
$|\lambda,\eta\rangle$ become the SU(2) generalized coherent states
$|j,\zeta_{0}\rangle$ with $\zeta_{0} = \tau^{\mp 1}$, respectively. 
Since $\eta$ is real and $|\eta|<1$, $\zeta_{0}$ is also real. Thus 
we have an intersection between the intelligent and coherent states.
The SU(2) coherent states which are simultaneously the 
$J_{2}$-$J_{3}$ intelligent states allow us to achieve the standard
noise limit (\ref{2.13}) due to the fact that $\zeta_{0}$ is real.
It means that the states in the coherent-intelligent intersection
lead to the best phase sensitivity among all the coherent states.
However, the standard noise limit can be surpassed by using the 
intelligent states which are not the generalized coherent states.

The decomposition of the intelligent states $|\lambda,\eta\rangle$
over the orthonormal basis is obtained by expanding the function 
$\Lambda(j,m_{0},\eta;\zeta)$ of Eq.\ (\ref{2.25}) into a Taylor 
series in $\zeta$. It is known \cite{GFun,Erd} that a function of 
the form (\ref{2.25}) is the generating function for the Lagrange 
polynomials:
\begin{equation}
\Lambda(j,m_{0},\eta;\zeta) = {\cal N}^{-1/2} \sum_{n=0}^{\infty} 
g_{n}^{(-j-m_{0},-j+m_{0})}(-1/\tau,-\tau) \zeta^{n} .  \label{2.29} 
\end{equation}
Actually, this series is finite, because we have 
\begin{equation}
g_{n}^{(-j-m_{0},-j+m_{0})} = 0 \;\;\;\; {\rm for} \; n>2j .
\label{2.30}      \end{equation}
The Lagrange polynomials are related to the Jacobi polynomials via
\cite{GFun}
\begin{equation}
g_{n}^{(\alpha,\beta)}(u,v) = (v-u)^{n} P_{n}^{(-\alpha-n,-\beta-n)}
\left( \frac{u+v}{u-v} \right) .   \label{2.31}
\end{equation}
Using this relation, we can write
\begin{equation}
|\lambda,\eta\rangle = {\cal N}^{-1/2} \sum_{m=-j}^{j}
\left[ \frac{(j+m)!(j-m)!}{(2j)!} \right]^{1/2} 
P_{j+m}^{(m_{0}-m,-m_{0}-m)}(x)\, t^{(j+m)/2}|j,m\rangle ,
\label{2.32}   \end{equation}
where we have defined
\begin{eqnarray}
& & x \equiv (1-\eta^{2})^{-1/2} , \nonumber \\
& & t \equiv 4(1-\eta^{2})/\eta^{2} = 4/(x^{2}-1) .   \label{2.33}
\end{eqnarray}
The normalization factor is
\begin{equation}
{\cal N} = \sum_{n=0}^{2j} \frac{n!(2j-n)!}{(2j)!}  
\left[P_{n}^{(j+m_{0}-n,j-m_{0}-n)}(x)\right]^{2} t^{n} .
\label{2.34}     \end{equation}
It follows from Eq.\ (\ref{2.30}) that the summation in (\ref{2.34})
can be continued up to infinity. Then, by using the summation 
theorem for the Jacobi polynomials \cite{GFun}, we find the closed
expression for the normalization factor:
\begin{equation}
{\cal N} = (-1)^{j-|m_{0}|} S_{+}^{j+m_{0}} S_{-}^{j-m_{0}} 
\frac{(j-m_{0})!(j+m_{0})!}{(2j)!} 
P_{j-|m_{0}|}^{(-2j-1,0)}\left( 1-\frac{2t}{S_{+}S_{-}} \right) ,
\label{2.35}    \end{equation}
where
\begin{equation}
S_{\pm} \equiv 1 + (x \pm 1)^{2} t/4 .  \label{2.36}
\end{equation}

The expression (\ref{2.34}) of ${\cal N}$ as a power series in $t$ 
is very convenient, because it enables us to write moments of the 
generator $J_{3}$ over the states $|\lambda,\eta\rangle$ as 
derivatives of ${\cal N}$ with respect to $t$. By using the property
$J_{3}|j,m\rangle = m|j,m\rangle$, we obtain
\begin{equation}
(\Delta J_{3})^{2} = \frac{t^{2}}{{\cal N}}
\frac{\partial^{2}{\cal N}}{\partial t^{2}} + \frac{t}{{\cal N}} 
\frac{\partial{\cal N}}{\partial t} - \left( \frac{t}{{\cal N}} 
\frac{\partial{\cal N}}{\partial t} \right)^{2} .    \label{2.37}
\end{equation}
By using the formula 
\begin{equation}
\frac{ d P_{n}^{(\alpha,\beta)}(x) }{dx} = \frac{ n+\alpha+\beta+1
}{ 2n }  P_{n-1}^{(\alpha+1,\beta+1)}(x)    \label{2.38}
\end{equation}
and the differential equation for the Jacobi polynomials,
we obtain the exact analytic expression for the
variance of $J_{3}$:
\begin{equation}
(\Delta J_{3})^{2} = \frac{\eta^{2}j}{2} \left[ 1 +
\frac{(j+|m_{0}|)}{j} (1-\eta^{2})
\frac{ P_{j-|m_{0}|-1}^{(1,-2j)}(1-2\eta^{2}) }{
P_{j-|m_{0}|}^{(0,-2j-1)}(1-2\eta^{2}) } \right] .  \label{2.39}
\end{equation}

\subsection{The phase sensitivity}
\label{ssec:su2_sens}

Substituting the above expression for $(\Delta J_{3})^{2}$ into Eq.\ 
(\ref{2.22}), we find the phase sensitivity of the interferometer 
fed with the SU(2) intelligent states:
\begin{equation}
(\delta \phi)^{2}_{{\rm int}} = \frac{G(j,m_{0},\eta)}{2j} ,
\label{2.40}   \end{equation}
where we have introduced the factor 
\begin{equation}
G(j,m_{0},\eta) \equiv \left[ 1 +\frac{(j+|m_{0}|) }{j} 
(1-\eta^{2}) \frac{ P_{j-|m_{0}|-1}^{(1,-2j)}(1-2\eta^{2}) }{
P_{j-|m_{0}|}^{(0,-2j-1)}(1-2\eta^{2}) }  \right]^{-1} .  
\label{2.41}     \end{equation}

In the case $m_{0}=\pm j$, i.e., for a state in the 
coherent-intelligent intersection, we have $G(j,m_{0},\eta)=1$, so
the phase sensitivity is at the standard noise limit $\delta \phi
= 1/\sqrt{N}$. However, the use of the SU(2) intelligent states that
do not belong to the coherent-intelligent intersection (i.e., with
$|m_{0}| \neq j$) can yield a considerable improvement of the 
measurement accuracy in comparison with the standard noise limit.
The quantitative measure of the improvement is the $G$-factor that
can be expressed as the ratio between the intelligent phase 
uncertainty and the standard noise limit:
\begin{equation}
G(j,m_{0},\eta) = (\delta \phi)^{2}_{{\rm int}}/
(\delta \phi)^{2}_{{\rm SNL}} .   \label{2.42}
\end{equation}
It follows from the properties of the Jacobi polynomials that in the
range considered here ($|\eta|<1$) we always have $G(j,m_{0},\eta)
\leq 1$, so the measurement accuracy is improved for SU(2) 
interferometers fed with intelligent light.

Numerical results are presented in Figs.\ 2 and 3. The function
$G(j,m_{0},\eta)$ is plotted in Fig.\ 2 versus $\eta$ for $j=15$ 
and various values of $m_{0}$. It is seen that for given $\eta$
the smaller the value of $m_{0}$, the smaller the $G$-factor.
We also see that the minimum value of $G(j,m_{0},\eta)$ (i.e., the 
best measurement accuracy) for given $j$ and $m_{0}$ is achieved 
when $\eta \rightarrow 0$. On the other hand, when 
$\eta \rightarrow 1$, the $G$-factor approaches unity. 
The phase sensitivity, i.e., the dependence of the minimum 
detectable phase shift $\delta\phi$ on the number $N=2j$ of photons
passing through the interferometer is illustrated in Fig.\ 3
where $\ln \delta\phi$ is shown as a function of $\ln N$ for 
$m_{0}=0$ and various values of $\eta$. 
It is seen that for a given value of $\eta$ the power law 
$\delta\phi \propto N^{-E}$ is a good approximation for large $N$.
In order to express formally the slope of the curves in Fig.\ 3
for large $N$, we introduce the exponent
\begin{equation}
E = - \left. \frac{ d (\ln \delta\phi) }{ d (\ln N) } 
\right|_{N \rightarrow \infty}  .  \label{Elim}
\end{equation}
This quantity is plotted versus $\eta$ in Fig.\ 4.
For $\eta \rightarrow 0$ the exponent $E$ approaches unity, which is 
the best available phase sensitivity. As $\eta$ increases, the 
exponent $E$ rapidly decreases to one half (the standard noise limit).

The dependence of the phase sensitivity on various parameters
can be further studied by considering limiting values of the 
$G$-factor. We start from the limit $\eta \rightarrow 1$. Putting 
$\varepsilon = 1-\eta^{2}$, we find, for $\varepsilon \ll 1$,
\begin{equation}
G(j,m_{0},\eta) \approx [1+2\varepsilon(j^{2}-m_{0}^{2})]^{-1} ,
\label{2.45}       \end{equation}
It means that for $\eta$ near $1$ the phase sensitivity approaches
the standard noise limit.
It is also not difficult to see that
\begin{equation}
\lim_{\eta \rightarrow 0} G(j,m_{0},\eta) = 
[1+(j^{2}-m_{0}^{2})/j]^{-1} ,           \label{2.43}
\end{equation}
and in this case we recover the approximate result of Hillery and 
Mlodinow \cite{HM}:
\begin{equation}
(\delta\phi)^{2}_{{\rm int}} \approx \frac{1}{ 2(j^2 - m_0^2 + j) } ,
\label{2.44}       \end{equation}
that holds for $\eta$ near zero. For $m_{0}=0$ the phase uncertainty
is minimized: 
\begin{equation}
(\delta\phi)_{{\rm int}} \approx \frac{1}{ \sqrt{2j(j+1)} }.
\end{equation} 
Because $j$ is just half the total number $N$ of photons passing 
through the interferometer, the phase sensitivity is of order $1/N$. 
We note that this sensitivity is achieved for the SU(2) intelligent
states without adding an active device to the interferometer.
Therefore, the total number of photons depends only on the value 
of $j$ for the input state. This allows us to avoid the duality in
the behavior of the phase sensitivity that occurs for the squeezed
input.

We also consider a subtlety that is concerned with the limit $\eta
\rightarrow 0$. It follows from the eigenvalue equation (\ref{2.21})
that for $\eta = 0$ the $J_2$-$J_3$ intelligent state 
$|\lambda,\eta\rangle$ transforms into the state $|j,m_0\rangle$ (an  
eigenstate of $J_3$ with eigenvalue $m_0$). However, this transition
should be treated with a great care, because it does not preserve
some basic properties of the intelligent states. In Eq.\ (\ref{2.22}),
that defines the phase sensitivity for the $J_2$-$J_3$ intelligent 
states, we have used the relation 
\begin{equation}
( \Delta J_2 )^2 = ( \Delta J_3 )^2 / \eta^2 .  \label{bprop}
\end{equation}
This property holds for any intelligent state with arbitrarily small
$|\eta|$. Therefore, we can take the limit $\eta \rightarrow 0$ for
the phase uncertainty $(\delta \phi)^{2}_{{\rm int}}$ calculated with 
the use of the relation (\ref{bprop}). But we see that the result
(\ref{2.44}) obtained in this way is quite different from the 
phase uncertainty (\ref{jmsens}) for the input states $|j,m\rangle$.
The reason for this discrepancy is that the ``intelligent'' relation
(\ref{bprop}) does not exist for the states $|j,m\rangle$. In other
words, the result depends on the order in which we use the relation 
(\ref{bprop}) and take the limit $\eta \rightarrow 0$. It means that 
the intelligent states $|\lambda,\eta\rangle$ with arbitrarily small 
$|\eta|$ and the states $|j,m_0\rangle$ may lead to different results.

This phenomenon arising for the SU(2) intelligent states in the limit
$\eta \rightarrow 0$ can be made more familiar if we recall a similar
situation that occurs for the canonical squeezed states. Consider two
canonically conjugate field quadratures, $Q = (a^{\dagger} + a)/2$ and 
$P = i(a^{\dagger} - a)/2$, which satisfy the uncertainty
relation $\Delta Q \Delta P \geq 1/4$. It is well known that this 
uncertainty relation is minimized by the canonical squeezed states
$|\zeta,\alpha\rangle$, which satisfy the eigenvalue equation
$(\eta Q + i P) |\zeta,\alpha\rangle = \lambda |\zeta,\alpha\rangle$.
Here $\alpha$ and $\zeta$ are displacement and squeezing amplitudes,
respectively, and $\eta = (1-\zeta)/(1+\zeta)$, $\lambda = (\alpha - 
\zeta \alpha^{\ast})/(1+\zeta)$. It is seen that the states 
$|\zeta,\alpha\rangle$ can be regarded as the intelligent states for
the Weyl-Heisenberg group. For instance, the relation 
$\Delta Q = \Delta P /|\eta|$ does hold for the $Q$-$P$ intelligent 
states with arbitrarily small values of $|\eta|$. However, for $\eta = 
0$ the $Q$-$P$ intelligent states transform into the eigenstates of 
the ``momentum'' operator $P$, and the above relation does not exist. 
Therefore, properties of the canonical squeezed states calculated using 
this relation may be different in the limit $\eta \rightarrow 0$ from 
corresponding properties of the momentum eigenstates.

\subsection{Quasi-intelligent states}
\label{ssec:su2_quasi}

The standard noise limit can also be surpassed by using two-mode 
states which are not exactly intelligent, but are close to optimizing
the uncertainty relation (\ref{2.19}). We will call such states
``quasi-intelligent.'' For example, we can imagine a state for which
the uncertainty product $( \Delta J_2 )^2 ( \Delta J_3 )^2$ is equal
to its minimum $\langle J_1 \rangle^2 /4$ times a numerical factor 
of order 1. For such a state we will get $(\delta\phi)^2 = \nu/
[ 4 ( \Delta J_2 )^2 ]$, where $\nu$ is the numerical factor.
If this state is squeezed in $J_3$ and swelled in $J_2$ (i.e.,
$\Delta J_2 \sim j$ for $j \gg 1$), then $\delta\phi$ will be of order 
$1/N$. An example of such a quasi-intelligent state was given by Yurke,
McCall and Klauder \cite{YMK} who considered the input state
$(|j,0\rangle + |j,1\rangle)/\sqrt{2}$. A simple calculation yields
\begin{eqnarray}
& & ( \Delta J_3 )^2 = \mbox{$\frac{1}{4}$} , \\
& & ( \Delta J_2 )^2 = \mbox{$\frac{1}{2}$} j(j+1) - 
\mbox{$\frac{1}{4}$} , \\
& & \langle J_1 \rangle = \mbox{$\frac{1}{2}$} [j(j+1)]^{1/2} .
\end{eqnarray}
We see that for $j \gg 1$ this state gives the uncertainty product
$( \Delta J_2 )^2 ( \Delta J_3 )^2 \approx j(j+1)/8 $ that 
is greater than its minimum $\langle J_1 \rangle^2 /4 = j(j+1)/16$
only by the factor $\nu = 2$. Then one obtains
\begin{equation}
(\delta\phi)^2 \approx \frac{1}{2 ( \Delta J_2 )^2 } \approx
\frac{1}{j(j+1)} .   
\end{equation}
Therefore the phase sensitivity is $\delta\phi \approx 2/N$ that 
differs from $(\delta\phi)_{{\rm int}} \approx \sqrt{2}/N$ only by 
the factor $\sqrt{2}$. This example shows that the optimization of 
the phase sensitivity is intimately related to the optimization of 
the uncertainty relation (i.e. the intelligence) and to the 
corresponding SU(2) squeezing.

\section{SU(1,1) interferometers with conventional input states}
\label{sec:su11_reg}

\subsection{The interferometer}
\label{ssec:su11_scheme}

In SU(1,1) interferometers four-wave mixers are employed instead
of beam splitters. The application of active optical devices, that 
do not preserve the total number of photons, makes it possible to 
achieve high measurement accuracy, especially when intelligent light
is used. On the other hand, this leads to the dual behavior of the
phase sensitivity, as in the case of the squeezed input.
Mathematical descriptions of SU(2) and SU(1,1) 
interferometers are rather similar, but the non-compactness of
the SU(1,1) Lie group leads to important physical distinctions
between interferometers employing passive and active devices.

An SU(1,1) interferometer is described schematically
in Fig.\ 5. Two light beams represented by mode annihilation 
operators $a_{1}$ and $a_{2}$ enter the input ports of the
first four-wave mixer FWM1. After leaving FWM1, the beams accumulate 
phase shifts $\phi_{1}$ and $\phi_{2}$, respectively, and then they 
enter the second four-wave mixer FWM2. The photons leaving the 
interferometer are counted by detectors D1 and D2.

For the analysis of such an interferometer it is convenient to 
consider the Hermitian operators
\begin{eqnarray}
& & K_{1} =\frac{1}{2}(a_{1}^{\dagger}a_{2}^{\dagger} 
+ a_{1}a_{2}) , \nonumber \\
& & K_{2} =\frac{1}{2i}(a_{1}^{\dagger}a_{2}^{\dagger} 
- a_{1}a_{2}) , \label{3.1}  \\
& & K_{3} =\frac{1}{2}(a_{1}^{\dagger}a_{1} 
+ a_{2}a_{2}^{\dagger}) \nonumber .
\end{eqnarray}
These operators form the two-mode boson realization of the SU(1,1)
Lie algebra:
\begin{eqnarray}
& &  [ K_{1}, K_{2} ] = -iK_{3}  , \nonumber \\   
& &  [ K_{2}, K_{3} ] = iK_{1}  , \label{3.2} \\
& &  [ K_{3}, K_{1} ] = iK_{2}  .  \nonumber 
\end{eqnarray}
It is also useful to introduce raising and lowering operators
\begin{eqnarray}
& & K_{+} = K_{1} + iK_{2} = a_{1}^{\dagger}a_{2}^{\dagger} , 
\nonumber \\
& & K_{-} = K_{1} - iK_{2} = a_{1}a_{2} . \label{3.3} 
\end{eqnarray}
The Casimir operator for any unitary irreducible representation is
a constant
\begin{equation}
K^{2} = K_{3}^{2} - K_{1}^{2} - K_{2}^{2} = k(k-1) . 
\label{3.4} \end{equation}
Thus a representation of SU(1,1) is determined by a single number
$k$ that is called the Bargmann index. For the discrete-series
representations \cite{SU11} the Bargmann index acquires discrete 
values $k=\frac{1}{2},1,\frac{3}{2},2,\ldots$.
By using the operators of Eq.\ (\ref{3.1}), one gets
\begin{equation}
K^{2} = \frac{1}{4}N_{d}^{2}
- \frac{1}{4} , \label{3.5}  \end{equation}
where
\begin{equation}
N_{d} = a_{1}^{\dagger}a_{1} - a_{2}^{\dagger}a_{2}   \label{3.6}
\end{equation}
is the photon-number difference between the modes. We see that
$N_{d}$ is an SU(1,1) invariant related to the Bargmann index $k$ 
via $k=\frac{1}{2}(N_{d}+1)$. The representation Hilbert space is 
spanned by the complete orthonormal basis $|k,n\rangle$ 
$(n=0,1,2,\ldots)$ that can be expressed in terms of Fock 
states of two modes:
\begin{equation}
|k,n\rangle = |n+2k-1\rangle_{1} |n\rangle_{2} .
\label{3.7}  \end{equation}

The actions of the interferometer elements on the vector
$\bbox{K}=(K_{1},K_{2},K_{3})$ can be represented as Lorentz boosts
and rotations in the (2+1)-dimensional space-time \cite{YMK}. 
FWM1 acts on $\bbox{K}$ as a Lorentz boost along the 
negative direction of the 2nd axis with 
the transformation matrix 
\begin{equation}
{\sf L}_{2}(-\beta) =  \left( \begin{array}{ccc} 
1 & 0 & 0 \\ 0 & \cosh\beta & - \sinh\beta \\
0 & - \sinh\beta & \cosh\beta   \end{array}  \right) . 
\label{3.8}            \end{equation}
As mentioned above, $\beta$ is related to the reflectivity $r$ of 
the four-wave mixer (when it is used as a phase-conjugating mirror) 
via $\sinh^{2} (\beta/2) = r$ \cite{ReWa}.
The transformation matrix of FWM2 is ${\sf L}_{2}(\beta)$, i.e., 
the two four-wave mixers perform boosts in opposite directions. 
Phase shifters rotate $\bbox{K}$ about the 3rd axis by an angle 
$\phi=-(\phi_{1}+\phi_{2})$. The transformation matrix of this 
rotation is ${\sf R}_{3}(\phi)$ of Eq.\ (\ref{2.9}).
The overall transformation performed on $\bbox{K}$ is
\begin{equation}
\bbox{K}_{{\rm out}} = 
{\sf L}_{2}(\beta){\sf R}_{3}(\phi){\sf L}_{2}(-\beta) \bbox{K} .
\label{3.9}  \end{equation}

The information on $\phi$ is once again inferred from the photon 
statistics of the output beams. One should measure the total number 
of photons in the two output modes, $N_{{\rm out}}$, or, 
equivalently, the operator $K_{3\,{\rm out}} = \frac{1}{2}
(N_{{\rm out}}+ 1)$. Fluctuations in $\langle K_{3\,{\rm out}}
\rangle$ restrict the accuracy of the phase measurement. The phase
uncertainty that determines the minimum detectable phase shift
is given by
\begin{equation}
(\delta\phi)^{2} = \frac{ (\Delta K_{3\,{\rm out}})^{2} }{ \left|
\partial\langle K_{3\,{\rm out}}\rangle /\partial\phi \right|^{2}} .
\label{3.10}  \end{equation}
From Eq.\ (\ref{3.9}), we find
\begin{equation}
K_{3\,{\rm out}} = (\sinh\beta \sin\phi) K_{1} 
+ \sinh\beta \cosh\beta (\cos\phi - 1) K_{2} 
+ (\cosh^{2}\!\beta - \sinh^{2}\!\beta \cos\phi) K_{3} .
\label{3.11}  \end{equation}

\subsection{The vacuum and coherent input states}
\label{ssec:su11_vac}

We consider here some typical cases when the input field is
prepared in the vacuum state, in the generalized coherent state 
and in the Glauber coherent state.
If only vacuum fluctuations enter the input ports, then Eq.\ 
(\ref{3.10}) with $K_{3\,{\rm out}}$ of Eq.\ (\ref{3.11}) reduces 
to the known result \cite{YMK}
\begin{equation}
(\delta\phi)^{2}_{\text{vac}} = \frac{ \sin^{2}\!\phi + 
\cosh^{2}\!\beta (1 - \cos\phi)^{2} }{ \sin^{2}\!\phi 
\sinh^{2}\!\beta } , \mbox{\hspace{0.6cm}} \phi \neq 0 .       
\label{3.12}  
\end{equation}
As $\phi \rightarrow 0$, these phase fluctuations are minimized, 
$(\delta\phi)^{2}_{\text{vac}} \rightarrow 1/\sinh^{2}\!\beta$.
We also consider a more complicated input state $|k,n\rangle
= |n+2k-1\rangle_1 |n\rangle_2$. The corresponding phase
uncertainty is obtained from Eq.\ (\ref{3.10}) by a straightforward
calculation:
\begin{equation}
(\delta\phi)^{2} = \frac{ \sin^{2}\!\phi + 
\cosh^{2}\!\beta (1 - \cos\phi)^{2} }{ \sin^{2}\!\phi 
\sinh^{2}\!\beta } \frac{k+n(2k+n)}{2(k+n)^2} ,
\mbox{\hspace{0.4cm}} \phi \neq 0 .     
\label{knsens1}  
\end{equation}
The vacuum state is obtained for $n=0$, $k=1/2$. Then Eq.\ 
(\ref{knsens1}) reduces to Eq.\ (\ref{3.12}). The phase uncertainty
(\ref{knsens1}) is minimized as $\phi \rightarrow 0$:
\begin{equation}
\lim_{\phi \rightarrow 0} (\delta\phi)^{2} = 
\frac{k+n(2k+n)}{2\sinh^{2}\!\beta (k+n)^2}.     
\label{knsens2}  
\end{equation}

In what follows we take for simplicity $\phi=0$, as in the SU(2) 
case. Once again, the experimenter can controll $\phi_{2}$ with a
feedback loop which maintains $\phi = -(\phi_{1}+\phi_{2}) = 0$
\cite{YMK}. Then Eq.\ (\ref{3.10}) with $K_{3\,{\rm out}}$ given 
by (\ref{3.11}) can be simplified to the form
\begin{equation}
(\delta\phi)^{2} = \frac{ (\Delta K_{3})^{2} }{ \sinh^{2}\!\beta
\langle K_{1} \rangle^{2} } , 
\mbox{\hspace{0.4cm}} \langle K_{1} \rangle \neq 0 .
\label{3.13}  \end{equation}

We next consider the SU(1,1) generalized coherent states.
These states are defined by \cite{Per}
\begin{eqnarray}
|k,\zeta\rangle & = & \exp(\xi K_{+} - \xi^{\ast} K_{-}) |k,0\rangle 
= (1-|\zeta|^{2})^{k} \exp(\zeta K_{+}) |k,0\rangle \nonumber \\
& = & (1-|\zeta|^{2})^{k} \sum_{n=0}^{\infty} \left[ \frac{
\Gamma(n+2k) }{ n!\Gamma(2k) } \right]^{1/2} \zeta^{n} |k,n\rangle ,
\label{3.14}     \end{eqnarray}
where $\zeta = (\xi/|\xi|)\tanh|\xi|$, so $|\zeta|<1$. In the case 
of the two-mode boson realization, the SU(1,1) coherent states can 
be recognized as the well-known two-mode squeezed states with $\xi$ 
being the squeezing parameter \cite{WE}. A simple calculation yields 
expectation values of the SU(1,1) generators over the 
$|k,\zeta\rangle$ coherent states \cite{WE}:
\begin{eqnarray}
& & (\Delta K_{3})^{2} = 2k|\zeta|^{2}/(1-|\zeta|^{2})^{2}  ,
\label{3.15}      \\ & & 
\langle K_{1} \rangle = 2k ({\rm Re}\, \zeta)/(1-|\zeta|^{2}) .
\label{3.16}     \end{eqnarray} 
Then Eq.\ (\ref{3.13}) reads 
\begin{equation}
(\delta\phi)^{2}_{{\rm coh}} = \frac{|\zeta|^{2}}{
2k\sinh^{2}\!\beta\, ({\rm Re}\,\zeta)^{2} } . 
\label{3.17}         \end{equation}
This phase uncertainty is minimized when $\zeta$ is real. Then 
one gets \cite{BBA:qso}
\begin{equation}
(\delta\phi)^{2}_{k,\beta} = \frac{1}{2k\sinh^{2}\beta} .
\label{3.18}         
\end{equation}
We see that this phase sensitivity depends only on the parameter 
$\beta$ of the four-wave mixer and on the photon-number difference
between the two input modes ($N_{d}=2k-1$). Therefore, $\zeta$ can
be taken to be zero, i.e., one can choose an input state with a 
fixed number of photons in the one mode and the vacuum in the other.
This is in accordance with the result (\ref{knsens2}) for the input 
state $|k,n\rangle$ with $n=0$.

The mean total number $\bar{N}$ of photons passing through the phase
shifters depends on both the input state and the four-wave mixer.
For the interferometer considered here, $\bar{N}$ is the total 
number of photons emitted by FWM1:
\begin{equation}
\bar{N} = 2 \langle K_{3}' \rangle - 1 ,   \label{Nm}             
\end{equation}
where $\bbox{K}' ={\sf L}_{2}(-\beta) \bbox{K}$, so we have
\begin{equation}
K_{3}' = (\cosh \beta)K_{3} - (\sinh\beta) K_{2} .  
\end{equation}
Calculating the expectation value for a coherent state with
real $\zeta$, we obtain
\begin{equation}
\bar{N} = 2k \frac{1+\zeta^2}{1-\zeta^2} \cosh\beta - 1 .
\end{equation}
Since $(\delta\phi)^{2}$ of Eq.\ (\ref{3.18}) is independent of 
$\zeta$, we may take $\zeta = 0$; then $\bar{N} = 2k\cosh\beta - 1$.
Once again, we have two ways for improving the measurement accuracy
of the interferometer: (i) by increasing the parameter $\beta$
of the four-wave mixer, or (ii) by increasing the photon-number 
difference $N_{d}=2k-1$ for the input state. In the first regime we
obtain the phase sensitivity for fixed input state 
($k = {\rm const}$):
\begin{equation}
\left. (\delta\phi)^{2} \right|_{k} = \frac{2k}{ (\bar{N}+1)^2
- (2k)^2 } .
\end{equation}
For $k=1/2$ ($N_{d}=0$), we recover the result for the vacuum input
\cite{YMK}:
\begin{equation}
(\delta\phi)^{2}_{{\rm vac}} = \frac{1}{ \bar{N} (\bar{N}+2) } .
\label{vsens}
\end{equation}
We see that the phase sensitivity approaches $1/N$. However, there 
is a problem with improvement of the measurement accuracy because 
the value of $\beta$ is restricted by properties of available 
four-wave mixers. On the other hand, the phase sensitivity for
fixed interferometer ($\beta = {\rm const}$) is
\begin{equation}
\left. (\delta\phi)^{2} \right|_{\beta} = \frac{ \cosh\beta }{
\sinh^2 \beta } \frac{1}{ (\bar{N}+1) }  .
\end{equation}
We see that the standard noise limit cannot be surpassed in this
regime. 

Next we consider the input state $|\alpha\rangle_1 |\alpha'\rangle_2$
where $|\alpha\rangle$ and $|\alpha'\rangle$ are the Glauber coherent
states. We easily find the following expectation values:
\begin{eqnarray}
& & \langle K_3 \rangle = ( |\alpha|^2 + |\alpha'|^2 + 1 )/2 , \\
& & ( \Delta K_3 )^2 = ( |\alpha|^2 + |\alpha'|^2 )/4 , \\
& & \langle K_1 \rangle = |\alpha| |\alpha'| \cos(\theta+\theta') , \\
& & \langle K_2 \rangle = |\alpha| |\alpha'| \sin(\theta+\theta') ,
\end{eqnarray}
where $\alpha = |\alpha|\, e^{i\theta}$, $\alpha' = |\alpha'|\, 
e^{i\theta'}$. For $\theta+\theta' = 0$ and $|\alpha| = |\alpha'|$,
we obtain
\begin{equation}
(\delta\phi)^{2}_{\alpha,\beta} = \frac{1}{2 |\alpha|^2 \sinh^2 \beta} ,
\end{equation}
\begin{equation}
\bar{N} = (2|\alpha|^2 + 1)\cosh\beta - 1 .
\end{equation}
These results are almost identical to Eqs.\ (\ref{vz1}) and (\ref{vz2})
for the SU(2) interferometer with squeezed input states; the only 
difference is the factor $2$ before $|\alpha|^2$. We again have two
regimes: (i) fixed input state ($\alpha = {\rm const}$)
and variable interferometer, or (ii) fixed interferometer
($\beta = {\rm const}$) and variable input state. The first regime
leads to the phase sensitivity of order $1/\bar{N}$, but is
technically more complicated.
The second regime is much more preferable from the technical point of 
view, but the phase sensitivity cannot be improved over the standard
noise limit. This duality in the behavior of the phase sensitivity
is a direct consequence of the fact that the SU(1,1) transformations
performed by the four-wave mixers do not preserve the total number of
photons.

\section{SU(1,1) interferometers with intelligent input states}
\label{sec:su11_int}

\subsection{The SU(1,1) intelligent states}
\label{ssec:su11_is}

We would like to surpass the standard noise limit by using the
SU(1,1) intelligent states. The commutation relation $[K_{2},K_{3}] 
= iK_{1}$ implies the uncertainty relation
\begin{equation}
(\Delta K_{2})^{2} (\Delta K_{3})^{2} \geq \frac{1}{4} 
\langle K_{1} \rangle^{2} .          \label{3.19}
\end{equation}
Therefore, Eq.\ (\ref{3.13}) can be written as
\begin{equation}
(\delta\phi)^{2}\geq\frac{1}{4\sinh^{2}\!\beta(\Delta K_{2})^{2}} .
\label{3.20}  \end{equation}
For intelligent states an equality is achieved in the uncertainty 
relation. Therefore, such $K_{2}$-$K_{3}$ intelligent states with 
large values of $\Delta K_{2}$ would allow us to measure small 
changes in $\phi$. The $K_{2}$-$K_{3}$ intelligent states 
$|\lambda,\eta\rangle$ are determined by the eigenvalue equation 
\begin{equation}
(\eta K_{2} + i K_{3}) |\lambda,\eta\rangle = \lambda 
|\lambda,\eta\rangle ,     \label{3.21}  
\end{equation}
where $\lambda$ is a complex eigenvalue and $\eta$ is a real 
parameter given by $|\eta| = \Delta K_{3}/ \Delta K_{2}$.
For $|\eta| > 1$, these states are squeezed in $K_{2}$, and for 
$|\eta| < 1$, they are squeezed in $K_{3}$. We consider here all 
the values of $\eta$. The scheme of Luis and Pe\v{r}ina \cite{LuisPer} 
can be used for producing both the SU(2) and the SU(1,1) intelligent
states. In particular, the $K_{2}$-$K_{3}$ intelligent states of Eq.\
(\ref{3.21}) can be generated in this scheme quite conveniently.
For these states, Eq.\ (\ref{3.20}) reads
\begin{equation}
(\delta\phi)^{2}_{{\rm int}} = \frac{1}{ 4\sinh^{2}\!\beta 
(\Delta K_{2})^{2} } 
= \frac{\eta^{2}}{ 4\sinh^{2}\!\beta (\Delta K_{3})^{2} } . 
\label{3.23}   \end{equation}

We will use the analytic representation in the basis of the SU(1,1) 
generalized coherent states $|k,\zeta\rangle$ \cite{Per}. This
basis is overcomplete, and any state in the Hilbert space can be 
expanded in it. For example, the SU(1,1) intelligent state
\begin{equation}
|\lambda,\eta\rangle = \sum_{n=0}^{\infty} C_{n} |k,n\rangle
\label{3.24}   \end{equation}
is represented by the function
\begin{equation}
\Lambda(k,\lambda,\eta;\zeta) = (1-|\zeta|^2)^{-k}\langle
k,\zeta^{\ast}|\lambda,\eta\rangle 
= \sum_{n=0}^{\infty} C_{n} \left[ \frac{
\Gamma(2k+n) }{ n!\Gamma(2k) } \right]^{1/2} \zeta^{n} ,
\label{3.25}   \end{equation}
which is analytic in the unit disk $|\zeta|<1$.
The analytic representation of the SU(1,1) intelligent states was 
studied in Ref.\ \cite{BBA:jpa}. 
The SU(1,1) generators act on $\Lambda(\zeta)$ as first-order 
differential operators \cite{Per}:
\begin{equation}
K_{+} = \zeta^{2} \frac{d}{d\zeta} + 2k\zeta ,  
\mbox{\hspace{0.5cm}} K_{-} = \frac{d}{d\zeta} ,
\mbox{\hspace{0.5cm}} K_{3} = \zeta \frac{d}{d\zeta} +k .    
\label{3.26}   \end{equation}
Then Eq.\ (\ref{3.21}) can be converted into a first-order linear 
homogeneous differential equation for $\Lambda(\zeta)$:
\begin{equation}
(\eta + 2\zeta - \eta \zeta^{2}) \frac{d\Lambda}{d\zeta} 
+ 2 (i\lambda +k -k\eta\zeta) \Lambda = 0 .     \label{3.27}
\end{equation}
The solution of this equation can be easily found to be
\begin{equation}
\Lambda(k,l,\eta;\zeta) = {\cal N}^{-1/2}
(1+\zeta/\tau)^{l}(1-\tau\zeta)^{-2k-l} ,
\label{3.28}    \end{equation}
where ${\cal N}$ is a normalization factor, and we have defined
\begin{eqnarray}
& & \tau \equiv \left( \sqrt{\eta^{2}+1} - 1 \right)/\eta ,
\;\;\;\;\;\;  |\tau| < 1 ,  \label{3.29} \\ 
& & \lambda(l) = i(k+l)\sqrt{\eta^{2}+1} .    \label{3.30}  
\end{eqnarray}
The analyticity condition for the function
$\Lambda(k,\lambda,\eta;\zeta)$ requires that $l$ can be only a 
positive integer or zero: $l=0,1,2,\ldots$. Then Eq.\ (\ref{3.30})
becomes a quantization condition which
means that the operator $\eta K_{2} + i K_{3}$ has a discrete 
spectrum, and the corresponding eigenstates and eigenvalues are
characterized by the quantum number $l$. 

In the simplest case $l=0$, the function $\Lambda(k,l,\eta;\zeta)$ 
represents the $K_{2}$-$K_{3}$ intelligent states which are 
simultaneously the SU(1,1) generalized coherent states 
$|k,\zeta_{0}\rangle$ with $\zeta_{0} = \tau$. Since $\eta$ is real,
$\zeta_{0}$ is also real.
Hence there is an intersection between the intelligent and 
coherent states. States which belong to this intersection allow us 
to achieve the measurement accuracy (\ref{3.18}) due to the fact 
that $\zeta_{0}$ is real. Therefore, these states lead to the best 
phase sensitivity among all the coherent states.
However, we will see that the noise level (\ref{3.18}) can be 
surpassed by using the SU(1,1) intelligent states which are not 
the generalized coherent states.

As in Sec.\ \ref{ssec:su2_is}, the function $\Lambda(k,l,\eta;\zeta)$ 
of Eq.\ (\ref{3.28}) is expanded into a Taylor series in $\zeta$
as the generating function for the Lagrange polynomials
\cite{GFun,Erd}: 
\begin{equation}
\Lambda(k,l,\eta;\zeta) = {\cal N}^{-1/2} \sum_{n=0}^{\infty} 
g_{n}^{(-l,2k+l)}(-1/\tau,\tau) \zeta^{n} .  \label{3.31} 
\end{equation}
By using the relation (\ref{2.31}) between the Lagrange and Jacobi 
polynomials, we obtain the decomposition of the SU(1,1) intelligent
sates over the orthonormal basis:
\begin{equation}
|\lambda,\eta\rangle = {\cal N}^{-1/2} \sum_{n=0}^{\infty}
\left[ \frac{n!\Gamma(2k)}{\Gamma(2k+n)} \right]^{1/2} 
P_{n}^{(l-n,-2k-l-n)}(x)\,  t^{n/2} |k,n\rangle ,        \label{3.32}
\end{equation}
where we have defined
\begin{eqnarray}
& & x \equiv (\eta^{2}+1)^{-1/2} , \nonumber \\
& & t \equiv 4(\eta^{2}+1)/\eta^{2} = 4/(1-x^{2}) .   \label{3.33}
\end{eqnarray}
By using the summation theorem for the Jacobi polynomials 
\cite{GFun}, we find the normalization factor:
\begin{equation}
{\cal N} = \sum_{n=0}^{\infty} \frac{n!\Gamma(2k)}{\Gamma(2k+n)}
\left[P_{n}^{(l-n,-2k-l-n)}(x)\right]^{2}  t^{n} 
= S_{+}^{l} S_{-}^{-2k-l} \frac{l!\Gamma(2k)}{\Gamma(2k+l)}
P_{l}^{(2k-1,0)}\!\left( 1+\frac{2t}{S_{+}S_{-}} \right) ,
\label{3.34}    \end{equation}
where
\begin{equation}
S_{\pm} \equiv 1 - (x \pm 1)^{2} t/4 .  \label{3.35}
\end{equation}

We can write moments of the generator $K_{3}$ over the states 
$|\lambda,\eta\rangle$ as derivatives of ${\cal N}$ with respect 
to $t$. By using the property $K_{3}|k,n\rangle = (k+n)|k,n\rangle$, 
we obtain
\begin{equation}
(\Delta K_{3})^{2} = \frac{t^{2}}{{\cal N}}
\frac{\partial^{2}{\cal N}}{\partial t^{2}} + \frac{t}{{\cal N}} 
\frac{\partial{\cal N}}{\partial t} - \left( \frac{t}{{\cal N}} 
\frac{\partial{\cal N}}{\partial t} \right)^{2} .     \label{3.36}
\end{equation}
By using formula (\ref{2.38}), we find the exact analytic expression 
for the variance of $K_{3}$:
\begin{equation}
(\Delta K_{3})^{2} = \frac{\eta^{2}k}{2} \left[ 1 +\frac{(2k+l)}{k} 
\frac{ (\eta^{2}+1) P_{l-1}^{(1,2k)}(2\eta^{2}+1) }{
P_{l}^{(0,2k-1)}(2\eta^{2}+1) } \right] .  \label{3.37}
\end{equation}

\subsection{The phase sensitivity}
\label{ssec:su11_sens}

Substituting the above result for $(\Delta K_{3})^{2}$ into Eq.\ 
(\ref{3.23}), we find the phase sensitivity of the interferometer 
fed with the SU(1,1) intelligent states:
\begin{equation}
(\delta \phi)^{2}_{{\rm int}} = \frac{G(k,l,\eta)}{2k 
\sinh^{2}\!\beta} ,
\label{3.38}   \end{equation}
where we have introduced the factor 
\begin{equation}
G(k,l,\eta) \equiv \left[ 1 + \frac{(2k+l)}{k} 
\frac{ (\eta^{2}+1) P_{l-1}^{(1,2k)}(2\eta^{2}+1) }{
P_{l}^{(0,2k-1)}(2\eta^{2}+1) } \right]^{-1} .  \label{3.39}
\end{equation}

In the case of the coherent-intelligent intersection, $l=0$, and
then $G(k,l,\eta) = 1$. Then the phase uncertainty is on the
noise level (\ref{3.18}).
The use of the SU(1,1) intelligent states that do not belong to the 
coherent-intelligent intersection (i.e., with $l\neq 0$) can
yield a great improvement of the measurement accuracy. The
quantitative measure of the improvement is the $G$-factor that
can be expressed as the ratio between the intelligent phase
uncertainty and the SU(1,1) coherent noise level (\ref{3.18}):
\begin{equation}
G(k,l,\eta) = \frac{ (\delta \phi)^{2}_{{\rm int}} }{
(\delta \phi)^{2}_{k,\beta} }
= 2k \frac{ (\delta \phi)^{2}_{{\rm int}} }{
(\delta \phi)^{2}_{\text{vac}} } .
\label{3.40}             \end{equation}
It follows from the properties of the Jacobi polynomials that the
$G$-factor is always less than unity, so the measurement accuracy
is improved for the SU(1,1) interferometers fed with intelligent
light. Quantitative results
are presented in Fig.\ 6, where the factor $G(k,l,\eta)$ is shown as 
a function of $\eta$ for $k=1/2$ and different values of $l$. We see 
that for given $\eta$ the larger the value of $l$, the smaller the 
$G$-factor. The best measurement accuracy for given 
$l$ and $k$ is achieved when $\eta \rightarrow 0$. For large values
of $\eta$, the $G$-factor approaches a limiting value. By using the
properties of the Jacobi polynomials, we find   
\begin{eqnarray}
& & \lim_{\eta\rightarrow 0} G(k,l,\eta) 
= \left[ 1 + l(2k+l)/k \right]^{-1}  , \label{3.41}   \\
& & \lim_{\eta\rightarrow \infty} G(k,l,\eta) = 
\left(1 + l/k \right)^{-1}  .
\label{3.42}  
\end{eqnarray}
An interesting property of SU(1,1) interferometers is that for 
$l \neq 0$ the coherent noise level (\ref{3.18}) is surpassed for any
value of $\eta$. The phase uncertainty (\ref{3.38}) for $\eta 
\rightarrow 0$ reads
\begin{equation}
\lim_{\eta\rightarrow 0} (\delta \phi)^{2}_{{\rm int}} =
\frac{1}{2 \sinh^{2}\!\beta [k+l(2k+l)]} .  \label{limuns}
\end{equation}

We have seen for the SU(2) interferometer that there is a subtlety 
concerned with the limit $\eta \rightarrow 0$. A similar problem also 
arises for the SU(1,1) interferometer. It follows from the eigenvalue 
equation (\ref{3.21}) that for $\eta = 0$ the $K_2$-$K_3$ intelligent 
state $|\lambda,\eta\rangle$ transforms into the state $|k,l\rangle$ 
(an eigenstate of $K_3$ with eigenvalue $k+l$). However, this 
transition does not preserve the relation 
\begin{equation}
( \Delta K_2 )^2 = ( \Delta K_3 )^2 / \eta^2 ,  \label{bprop11}
\end{equation}
which has been used in Eq.\ (\ref{3.23}) that defines the phase 
sensitivity for the intelligent states. The property (\ref{bprop11}) 
holds for any intelligent state with arbitrarily small $|\eta|$. 
Therefore, we can take the limit $\eta \rightarrow 0$ for the phase 
uncertainty $(\delta \phi)^{2}_{{\rm int}}$ calculated with the use of 
the relation (\ref{bprop11}). But we see that the result (\ref{limuns}) 
obtained in this way is different from the result (\ref{knsens2}) for 
the input states $|k,n\rangle$. This discrepancy occurs because the 
``intelligent'' relation (\ref{bprop11}) does not exist for the states 
$|k,n\rangle$. Therefore, the intelligent states $|\lambda,\eta\rangle$ 
with arbitrarily small $|\eta|$ and the states $|k,l\rangle$ lead to 
different phase uncertainties.

We proceed by examining the phase sensitivity $\delta\phi(\bar{N})$
for the intelligent input. The mean total number $\bar{N}$ of photons
passing through the phase shifters is given by Eq.\ (\ref{Nm}).
It follows directly from the eigenvalue equation (\ref{3.21}) that
$\langle K_{2} \rangle = ({\rm Re}\, \lambda)/\eta$, 
$\langle K_{3} \rangle = {\rm Im}\, \lambda$. Then we use the 
quantization condition (\ref{3.30}) and find
\begin{equation}
\langle K_{2} \rangle = 0, \;\;\;\;\; 
\langle K_{3} \rangle = (k+l)\sqrt{\eta^{2}+1} .   \label{3.45}
\end{equation}
Then $\bar{N}$ is given by
\begin{equation}
\bar{N} = 2 \cosh\beta\, (k+l)\sqrt{\eta^{2}+1} - 1 .   
\label{3.46}
\end{equation}
We see that $\bar{N}$ depends on the parameters $k$, $l$, $\eta$ of
the input state and on the parameter $\beta$ of the interferometer.
The phase sensitivity for fixed input state is
\begin{equation}
\left. (\delta \phi)^{2} \right|_{k,l,\eta} = \frac{1}{k} 
\frac{ 2(k+l)^{2} (\eta^{2}+1) G(k,l,\eta) }{ (\bar{N}+1)^{2} 
- 4(k+l)^{2} (\eta^{2}+1)} .         \label{3.47}     
\end{equation}
For $\eta \rightarrow 0$ this phase sensitivity is
\begin{equation}
\left. (\delta \phi)^{2} \right|_{k,l} = \frac{2(k+l)^2}{
(l^2 + 2k l + k) } \frac{1}{ (\bar{N}+1)^2 - 4(k+l)^2 } .
\end{equation}
For $l=0$ and $k=1/2$, this result reduces to Eq.\ (\ref{vsens})
for the vacuum input. For $k=1$, we obtain
\begin{equation}
\left. (\delta \phi)^{2} \right|_{l} = 
\frac{2}{ (\bar{N}+1)^2 - 4(l+1)^2 } .
\end{equation}
We see that this regime can yield a phase sensitivity of order 
$1/\bar{N}$, that, of course, depends on the available range of 
$\beta$. 

In the regime of fixed interferometer ($\beta = {\rm const}$)
and variable input state, the phase sensitivity depends on the 
three parameters of the state: $k$, $l$ and $\eta$.
We study the dependence $\delta\phi(\bar{N})$ numerically:
for fixed $\sinh^2 \beta = 1$ and some values of $k$ and $\eta$,
we evaluate numerically $(\delta \phi)^{2}$ of Eq.\ (\ref{3.38})
and $\bar{N}$ of Eq.\ (\ref{3.46}) for $l=1,2,\ldots,150$. These 
results are presented in Fig.\ 7 where $\ln \delta\phi$ is plotted 
versus $\ln \bar{N}$ for $k=1/2$ and various values of $\eta$. 
In the region of large $\bar{N}$ (small phase uncertainty), a good 
approximation is the power law $\delta\phi \propto \bar{N}^{-E}$. 
We introduce the exponent 
\begin{equation}
E = - \left. \frac{ d (\ln \delta\phi) }{ d (\ln \bar{N}) } 
\right|_{\bar{N} \rightarrow \infty} , \label{Elim11}
\end{equation}
which expresses the slope of the curves in Fig. 7 for large $\bar{N}$.
This quantity is plotted in Fig.\ 8 versus $\eta$. It is seen that 
$E$ approaches unity for $\eta \rightarrow 0$ and rapidly decreases to 
one half as $\eta$ increases. 

Using the limit (\ref{3.41}), we find the phase sensitivity for 
fixed interferometer ($\beta = {\rm const}$) and the input
state with $\eta \rightarrow 0$, fixed $k$ and variable $l$:
\begin{equation}
\left. (\delta \phi)^{2} \right|_{k,\beta} \approx  
\frac{ 2\coth^2 \beta }{ (\bar{N}+1)^2 - 4(k^2 -k)\cosh^2 \beta } .
\end{equation}
This phase sensitivity is optimized for $k=1/2$:
\begin{equation}
\left. (\delta \phi)^{2} \right|_{\beta} \approx  
\frac{ 2\coth^2 \beta }{ (\bar{N}+1)^2 + \cosh^2 \beta} .
\end{equation}
Thus we see that the interferometer operated in the regime of fixed 
$\beta$ can achieve a phase sensitivity of order $1/\bar{N}$. It 
means that SU(1,1) interferometers with intelligent input states can 
surpass the standard noise limit in the both regimes: for fixed input 
state, and for fixed interferometer. This remarkable property 
distinguishes the intelligent states from the other states discussed 
above.

\section{Discussion and conclusions}
\label{sec:conc}

In this paper we considered in detail the phase sensitivity of passive 
and active interferometers characterized by the SU(2) and SU(1,1) 
groups respectively, for various types of input states. A usual method 
to reduce the quantum noise in interferometers is by the application 
of squeezing-producing active devices. We showed that the use of such 
active devices (e.g., four-wave mixers) leads to a duality in the 
behavior of the phase sensitivity. If the total number $\bar{N}$ of 
photons passing through the phase shifters is determined by means of 
the squeezing parameter $\beta$ of the four-wave mixer, the 
interferometer can achieve the phase sensitivity 
$\delta\phi \sim 1/\bar{N}$. However, if $\bar{N}$ is determined by 
changing parameters of an input state (e.g., the intensity of a 
coherent laser beam), the phase sensitivity cannot generally surpass 
the standard noise limit $\delta\phi \sim 1/\bar{N}^{1/2}$. We showed 
that these limitations can be overcome by the use of the remarkable 
squeezing properties of the intelligent states. On the one hand, 
the SU(2) intelligent states can lead to the phase sensitivity 
$\delta\phi \sim 1/\bar{N}$ in SU(2) interferometers without 
additional squeezing-producing devices. This avoids the duality 
mentioned above. On the other hand, SU(1,1) interferometers fed with 
the SU(1,1) intelligent states achieve phase sensitivity of order 
$1/\bar{N}$ in both regimes: for variable squeezing parameter 
$\beta$ of the four-wave mixer, and for variable intensity of the 
input state.

The quantum noise is formally expressed via the uncertainty relations. 
Quantum states which optimize the uncertainty relations lead to  
minimum noise. This property is called intelligence and it can be 
manifested in arbitrarily strong squeezing achieved by the intelligent 
states. It means that while the uncertainty of a quantum observable is 
dramatically reduced, the uncertainty of a conjugate observable 
is increased as little as allowed by quantum theory. The quantum noise 
in SU(2) and SU(1,1) interferometers is expressed via the uncertainty 
relations for the Hermitian generators of the correponding groups. 
Therefore, the best way to reduce this noise is by using the SU(2) and 
SU(1,1) intelligent states respectively, which are highly squeezed for 
an appropriate group generator ($J_3$ and $K_3$ respectively, in the 
schemes considered here). The production of these states by means of 
advanced experimental techniques looks quite realistic in the near 
future. We also note that the powerful analytic method used for 
calculations with the intelligent states can be of considerable 
interest to workers in quantum optics.

In the present paper we adopted an ideal assumption that the input
two-mode state has a definite total number of photons $N=2j$ [for 
an SU(2) interferometer] or a definite photon-number difference 
$N_{d}=2k-1$ [for an SU(1,1) interferometer]. In other words, we
considered input states belonging to irreducible representations
of SU(2) and SU(1,1). A more realistic assumption should deal with
an input states which is a superposition of the intelligent states
with different values of $j$ or $k$. Properties of such a 
superposition state will depend on the photon-number sum and 
difference distribution in the SU(2) and the SU(1,1) case, 
respectively.

\acknowledgements

C.B. gratefully acknowledges the financial help from the Technion.
A.M. was supported by the Fund for Promotion of Research
at the Technion, by the Technion -- VPR Fund, and by GIF --- 
German-Israeli Foundation for Research and Development.
We thank the Referees for valuable comments.

\begin{figure}
\caption{An SU(2) interferometer. Two light modes $a_{1}$ and
$a_{2}$ are mixed by beam splitter BS1, accumulate phase shifts
$\phi_{1}$ and $\phi_{2}$, respectively, and then they are again
mixed by beam splitter BS2. The photons in the output modes are
counted by detectors D1 and D2.}
\end{figure}
\begin{figure}
\caption{The factor $G(j,m_{0},\eta)$ of Eq.\ (\protect\ref{2.41})
versus $\eta$ for $j=15$ and various values of $m_{0}$.}
\end{figure}
\begin{figure}
\caption{$\ln \delta\phi$ as a function of $\ln N$ for an SU(2)
interferometer using the SU(2) intelligent states with $m_{0}=0$ 
and various values of $\eta$.}
\end{figure}
\begin{figure}
\caption{The exponent $E$ of Eq.\ (\protect\ref{Elim}) versus $\eta$ 
for an SU(2) interferometer using the SU(2) intelligent states with 
$m_{0}=0$.}
\end{figure}
\begin{figure}
\caption{An SU(1,1) interferometer. Two light modes $a_{1}$ and
$a_{2}$ are mixed by four-wave mixer FWM1, accumulate phase shifts
$\phi_{1}$ and $\phi_{2}$, respectively, and then are again
mixed by four-wave mixer FWM2. The photons in the output modes are
counted by detectors D1 and D2.}
\end{figure}
\begin{figure}
\caption{The factor $G(k,l,\eta)$ of Eq.\ (\protect\ref{3.39}) 
versus $\eta$ for $k=1/2$ and different values of $l$.}
\end{figure}
\begin{figure}
\caption{$\ln \delta\phi$ as a function of $\ln \bar{N}$ for an 
SU(1,1) interferometer with $\sinh^2 \beta = 1$,
using the SU(1,1) intelligent states with $k=1/2$ 
and various values of $\eta$. The values of $\delta\phi$ and 
$\bar{N}$ are calculated for $l=1,2,\ldots,150$.}
\end{figure}
\begin{figure}
\caption{The exponent $E$ of Eq.\ (\protect\ref{Elim11}) versus $\eta$ 
for an SU(1,1) interferometer with $\sinh^2 \beta = 1$, using the 
SU(1,1) intelligent states with $k=1/2$.}
\end{figure}

\end{document}